# Multi-step LRU: SIMD-based Cache Replacement for Lower Overhead and Higher Precision

Hiroshi Inoue
IBM Research - Tokyo
Tokyo, Japan
inouehrs@jp.ibm.com

*Abstract*—A key-value cache is a key component of many services to provide low-latency and high-throughput data accesses to a huge amount of data. To improve the end-to-end performance of such services, a key-value cache must achieve a high cache hit ratio with high throughput. In this paper, we propose a new cache replacement algorithm, *multi-step LRU,* which achieves high throughput by efficiently exploiting SIMD instructions without using per-item additional memory (*LRU metadata*) to record information such as the last access timestamp. For a small set of items that can fit within a vector register, SIMD-based LRU management without LRU metadata is known (*in-vector LRU*). It remembers the access history by reordering items in one vector using vector shuffle instruction. In-vector LRU alone cannot be used for a caching system since it can manage only few items. Set-associative cache is a straightforward way to build a large cache using in-vector LRU as a building block. However, a naive set-associative cache based on in-vector LRU has a poorer cache hit ratio than the original LRU although it can achieve a high throughput. Our multi-step LRU enhances naive set-associative cache based on in-vector LRU for improving cache accuracy by taking both access frequency and access recency of items into account while keeping the efficiency by SIMD instructions. Our results indicate that multi-step LRU outperforms the original LRU and GCLOCK algorithms in terms of both execution speed and cache hit ratio. Multi-step LRU improves the cache hit ratios over the original LRU by implicitly taking access frequency of items as well as access recency into account. The cache hit ratios of multi-step LRU are similar to those of ARC, which achieves a higher a cache hit ratio in a tradeoff for using more LRU metadata.

*Keywords—Cache replacement, LRU, SIMD*

## I. INTRODUCTION

To provide low-latency and high-throughput accesses for a huge amount of data, key-value caches, such as Memcached [1] and Redis [2], play a critically important role in many of today's services. Such key-value caches typically use least recently used (LRU) or its approximation (pseudo-LRU) as a replacement algorithm. For example, Memcached maintains a doubly linked list for key-value items in each size class to remember the access history. Redis uses a pseudo-LRU; it records the last access timestamp for each item and evicts the LRU item among three randomly selected items. Fan *et al.* [3] showed that using a CLOCK-based pseudo-LRU, which uses only one bit per item as LRU metadata to show the item is recently used, in combination with other improvements, can significantly enhance the overall performance of Memcached. LRU based on a doubly link list uses two pointers (16 bytes for 64-bit systems) for each item to maintain the information necessary for replacement. This memory overhead due to LRU metadata is not negligible, especially when the average size of cached items is small and the number of items becomes huge, as is often observed in real-world services. If we can reduce the memory overhead for LRU metadata, we can cache more items and improve the cache hit ratio. Hence pseudo-LRU policies often aim to reduce memory overhead due to per-item LRU metadata. For example, CLOCK uses only one bit per item for LRU metadata. Also, the overhead in CPU time for maintaining the linked list is not negligible for achieving high throughput. Pseudo-LRU algorithms also tend to reduce the CPU time for cache replacement compared with the exact LRU.

If the size of each item to cache is large, e.g., as in an operating system's memory subsystems that manage memory pages, it is reasonable to use more LRU metadata for achieving higher cache hit ratios compared with the LRU by considering access frequency as well as access recency [4, 5]. These algorithms typically use additional LRU metadata to remember the recently evicted items. For example, Adaptive Replacement Cache (ARC) [5] keeps two separated linked lists: one for items used only once and the other for items accessed twice or more. ARC keeps keys (page IDs) in two linked lists twice as many as the memory pages the system can cache. Such an approach can be rationalized for caching an object that is much larger than the key. However, such overheads may be excessive for caching a huge number of small objects, although the improved accuracy is attractive for any caching system.

In this paper[1], we propose a cache replacement algorithm called *multi-step LRU* that can achieve lower runtime overheads in both memory and CPU while achieving higher precision. Multi-step LRU 1) does not use per-item LRU metadata for maintaining access history, 2) can efficiently exploit SIMD (vector) instructions, and 3) improves the cache hit ratio compared to LRU by prioritizing frequently accessed items as well as recently accessed items. In this paper, we assume that the key-value cache is used to maintain key-value pairs, each of which consists of a key and a pointer to the cached object assigned to the key. Hence, we assume the size of both key and value are fixed to 64 bits regardless of the actual size of the objects to cache. Figure 1 shows an example of expected use

---

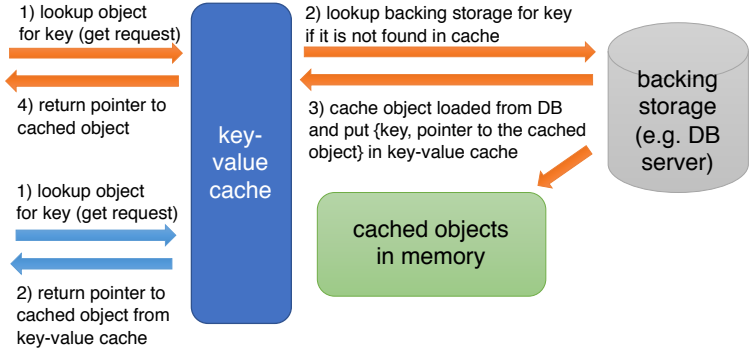
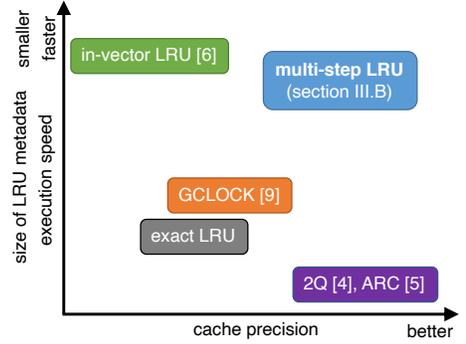

Figure 1. Expected use cases of multi-step LRU for key-value cache that manages pointer to cached objects. Blue arrows show query that hits in cache and orange arrows show query that does not hit cache.

Figure 2. Execution speed and precision for various replacement algorithms. Our multi-step LRU achieves both high speed and high precision.

cases with a query that hits in the cache (blue arrows) and another query that does not (orange arrows).

First, we briefly explain SIMD-based LRU management by Wang et al. [6], which we refer as *in-vector LRU* in this paper. It is an efficient implementation to do the exact LRU replacement for a small number of items that can fit in one vector register without using per-item LRU metadata. This is a building block of our multi-step LRU algorithm. In-vector LRU remembers the access history information by ordering items in one vector from the most recently used (MRU) item to the LRU item. Keeping items in order of recent accesses is generally too costly to do with a large number of items. However, for a small number of items that can fit within one vector register, we can efficiently reorder items by using permute (shuffle) instructions, which can rearrange items in one vector.

In-vector LRU alone cannot be used for a caching system since it can manage only few or several items. We can build a large caching system based on in-vector LRU for $P$ items by using a $P$-way set-associative cache. Here, $P$ is the number of items that one SIMD instruction can handle, e.g., $P = 4$ for 64-bit keys and Intel's AVX instruction set, which uses 256-bit vector registers. In a $P$-way set-associative cache, each item is assigned to a set on the basis of the hash value of the key. Each set consists of $P$ items; hence, in-vector LRU can be used to find the item to evict when adding a new item to the set that is already full. Although this $P$-way set-associative cache is quite efficient in terms of processing performance, it suffers from more cache misses than LRU or pseudo-LRU algorithms.

We propose *multi-step LRU* for improving the cache hit ratio while maintaining the efficiency of in-vector LRU, proposed multi-step LRU approximates LRU to manage replacement within each set consisting of more than $P$ items. We assume each set consists of $M \times P$ items (or $M$ vectors). The first vector contains MRU items, and the $M$-th vector contains LRU items. Within a vector, items are also ordered from the MRU to LRU ones. Hence, the last item of the $M$-th vector is the LRU item in the set and to be evicted next. With multi-step LRU, a newly added item is placed at the first (MRU) position of the $M$-th vector, instead of the MRU position of the first vector. Since we add the new item in the last vector and evict the LRU item from the same vector, we need to update only this $M$-th vector, whereas exact LRU requires updating all $M$ vectors. For a get request, we scan all vectors to find the requested item. If an item is found in the $i$-th vector, for example, we move this item to the MRU position of the $i$-th vector instead of the MRU position of the first vector to avoid updating multiple vectors. If the found item is already at the MRU position on the $i$-th vector, we swap this item with the LRU item of the ($i$-1)-th vector. Hence, a newly added item needs to be frequently accessed to be placed in the first vector, whereas any item can go directly to the MRU position of the entire set with only one access in exact LRU. Multi-step LRU is efficient because we only need to modify one vector or swap items between two vectors for both put and get operations. Moreover, with this multi-step LRU only frequently accessed items are placed in vectors other than the last one; items in the non-last vectors are not evicted from a cache by an item that is accessed only once. This characteristic makes it possible to protect frequently accessed items from being evicted by items, each of which is accessed only once without explicitly counting the number of accesses for each item. Hence, this very simple and efficient algorithm inherently prioritizes frequently accessed items over items accessed only once without explicitly managing them separately, as with ARC.

Figure 2 summarizes the tradeoffs between cache precision and execution speed for various replacement algorithms.

## II. RELATED WORK

### A. Cache Replacement Algorithms

Because of its importance, cache replacement has been studied for various purposes such as database management systems, networking software, and storage systems. Therefore, many replacement algorithms have been proposed. Simple replacement algorithms include random, FIFO (first-in first-out), LFU (least frequently used), and LRU.

To improve these simple algorithms in both runtime efficiency and cache precision, many enhanced variants have been developed and widely used. For example, CLOCK is one of the most well-known LRU variants. CLOCK is much more efficient at runtime than the LRU since it uses a global iterator (a hand) and a per-item one-bit flag to show the recent use as LRU metadata. To determine the item to evict, it iterates items and evicts the first item without the flag. There are many algorithms that further enhance CLOCK, such as CAR [7], CLOCK-Pro [8] and generalized CLOCK (GCLOCK) [9]. GCLOCK is a widely used algorithm to improve the cache hit ratio by using a reference count instead of one-bit flag. The reference count is incremented for each get request to the item. For eviction, GCLOCK iterates items and decrements the reference count until it finds an item whose reference count is zero. By using a multi-bit counter instead of a one-bit flag, GCLOCK can avoid evicting frequently used items as a tradeoff for additional memory consumption for LRU metadata. Our multi-step LRU are more memory efficient than CLOCK-based algorithms since we do not need any per-item LRU metadata. In this study, we use GCLOCK as a baseline for performance comparisons since it is one of the most representative of the CLOCK variants.

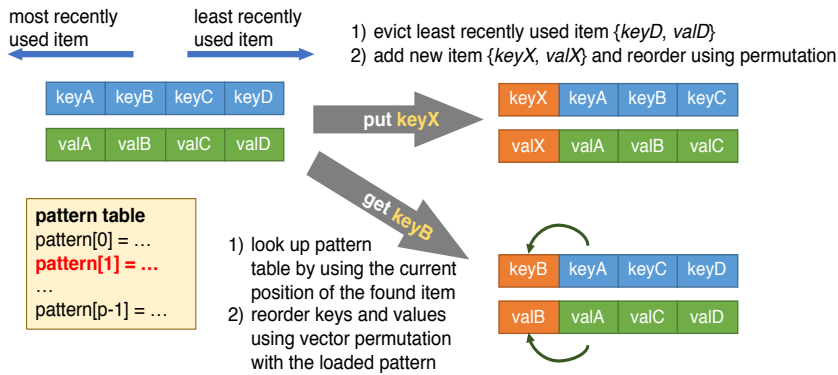

Figure 3. Overview of in-vector LRU with four items ($P = 4$) for the put and get operations. Four keys are stored in a contiguous memory region and reordered in order of recent uses. Associated values are also rearranged in the same order.

LRU-2 [10] is a well-known LRU variant, which can improve the cache hit ratio. LRU-2 (or LRU-k in general) determines the item to evict based on the second ($k$-th) most recent access instead of the most recent access, which the original LRU uses. The algorithms 2Q [4], MQ [11], and ARC [5] improve upon LRU-2 in both implementation complexity and cache precision. Our multi-step LRU can achieve similar gains in the improved cache hit ratio compared with the original LRU, as with ARC, while the computation overheads in memory and CPU are much smaller with ours. In experiments, we evaluate ARC as a replacement algorithm that yields the highest precision and compared it with our multi-step LRU.

*B. Set-associative Cache*

For implementing cache replacement in hardware for a processor's cache memory, a set-associative cache is often used to reduce the hardware implementation cost [12]. In a hardware cache memory, each key-value item to be cached consists of the memory address as the key and a small chunk (e.g. 64 byte) of data stored in the memory as the value.

In a set-associative cache whose size is $N$ items in total, these items are divided into equal-sized sets, and each item is assigned to one of the sets on the basis of the hash value of the key. When the size of a set is $P$ items, the cache is called a $P$-way set-associative cache. Cache replacement is done within each set; when inserting a new item into a $P$-way set-associative cache, one of the existing $P$ items is evicted to make room for the new item. Hence, if we have a cache replacement algorithm for $P$-items, we can build a large set-associative cache using the replacement algorithm by assigning each key onto a set based on the hash value of the key. Since only $P$ items in the same set are the candidates for eviction, the cost for implementing cache replacement is drastically reduced by using small $P$ in a tradeoff for increased cache misses. If one set becomes full and a new item comes into this set, one of the existing items must be evicted, even if there are unused spaces in other sets. Using smaller $P$ increases the risk of such cache misses (called conflict miss). When multiple hot items are assigned to the same set, they make each other frequently evicted, which may increase the overall number of cache misses. When $P = N$, i.e., the entire cache consists of only one huge set, the cache is called a *fully associative cache*. When $P = 1$, the cache is called a *direct-mapped cache*. Hardware designers need to decide the associativity between the fully associative cache and direct-mapped cache based on the tradeoff of hardware cost and hit ratio. On today's high-performance processors, the cache memory implementation typically uses a 2-way to 16-way set-associative cache to balance implementation cost and performance. Within each set, a hardware-friendly pseudo-LRU algorithm similar to CLOCK is often used. In this paper, we used a set-associative cache implemented in software to efficiently exploit the vector registers, the size of which is limited by hardware.

III. OUR REPLACEMENT ALGORITHM

In this section, we present our cache replacement algorithms. First, we briefly explain in-vector LRU [6], an algorithm to implement exact (i.e. not approximate) LRU for a small number ($P$) of key-value items without per-item LRU metadata. Then we explain a set-associative cache with our proposed multi-step LRU for constructing a large cache using in-vector LRU.

*A. In-vector LRU for Small Set*

We explain in-vector LRU [6] for four items ($P = 4$), e.g., items with a 64-bit key using 256-bit vector registers, although the algorithm is not limited to this specific configuration. We tested multi-step LRU with $P = 8$ as well as $P = 4$ in the evaluations.

Figure 3 illustrates an overview of the put and get operations. We also show a pseudo-code for these operations using Intel's intrinsics [13] in supplemental material. We consider a case having $P$ key-value items in memory; $P$ keys are stored in a contiguous memory region to efficiently exploit SIMD instructions. $P$ values are also stored in another contiguous region. These $P$ keys (and associated $P$ values) are sorted; the first key in a vector is the MRU one and the last ($P$-th) key is the LRU one, as shown in Figure 3.

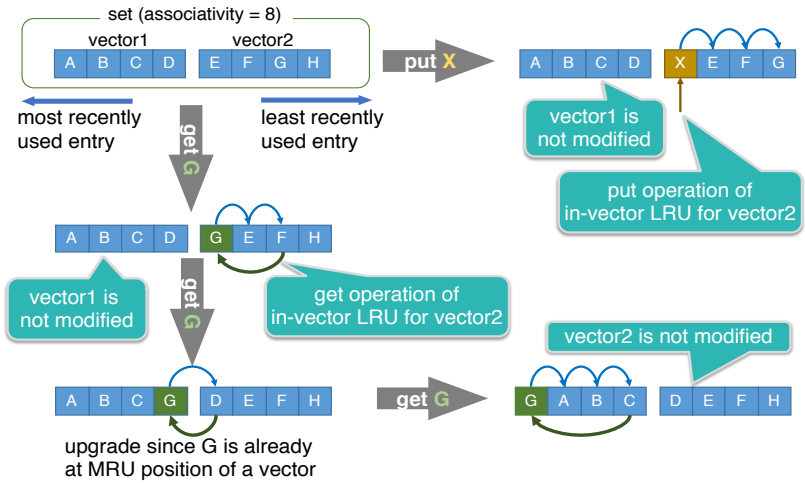

Figure 4. Overview of multi-step LRU with $M = 2$ and $P = 4$ for put and get operations. Only keys are depicted for simplicity; values associated with keys also move along with keys.

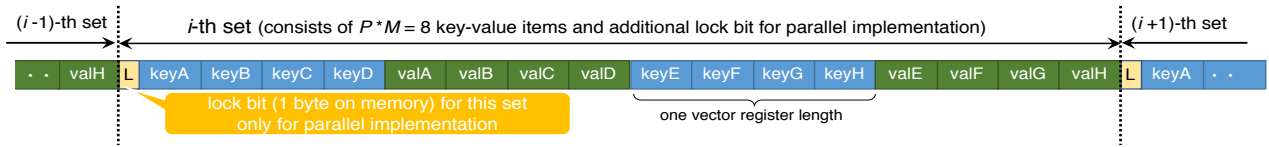

Figure 5. Overview of multi-step LRU with $M = 2$ and $P = 4$ for put and get operations. Only keys are depicted for simplicity; values associated with keys also move along with keys.

When a get query arrives, a get operation checks all $P$ keys by using SIMD compare instructions. If it hits any of the keys, the get operation returns the associated value then rearranges keys and values to move this key-value item to the first (MRU) position. For this, we use a vector permute instruction for rearranging elements within a vector. A get operation consists of the following steps:

1) Load $P$ keys from memory into a vector register.
2) Check $P$ keys against the query using SIMD compare.
3) If no key hits, the operation ends as a cache miss.
4) When the $i$-th key hits, look up the in-memory constant table of permutation patterns using $i$ as the index.
5) Move the $i$-th key into the MRU position by a permutation instruction with the pattern loaded in step 4 and store the rearranged keys in the vector register into the original memory location.
6) Load $P$ values into a vector register, rearrange them using the same pattern, and store the values back into memory.
7) Return the result (a cache hit).

To use in step 4, we create a small constant table (*patternTable* in pseudo code) in memory to determine the permutation pattern efficiently.

A put operation simply evicts the LRU item if the cache is full and inserts the new item. Since the last ($P$-th) item is always the LRU item, we do not need to do anything for finding the item to evict. The put operation can be simply implemented as follows (assuming the cache is full and cache miss is already confirmed).

1) Load $P$ keys into a vector register.
2) Move the LRU key into the MRU position using permutation.
3) Replace the LRU key with the new key then store them back into memory.
4) Do steps 1 to 3 for values.

Before these steps, we check for an empty slot for all vectors. If there is an empty slot, we just put the new item there.

Since rearranging items by vector load, permute, and store instructions is much more efficient than rearranging items by scalar load and stores, it does not incur significant overhead at each get and put operation. The put operation is much simpler than in other algorithms since the item to evict is always placed at the last position in the vector, and we do not need to scan items or access the linked list to find the LRU item.

We use the vector permute instructions based on a constant value fetched from an in-memory constant table to rearrange items within a vector. The idea of using a constant table to feed the pattern for the permutation instruction is not new. Similar techniques using a constant table have been used for decoding compressed information such as UTF-8 characters [14] or compressed posting lists [15].

Although in-vector LRU is quite efficient in execution performance, the number of items that can be managed by it is limited by the width of a vector register; hence, we cannot expect that in-vector LRU alone can be used for realistic caching systems even considering the increasing size of the vector register on the latest processors.

*B. Set-associative Cache with Multi-step LRU*

A set-associative cache, which is often used to implement cache memory of processor hardware, is a straight-forward choice to implement a caching system for a large number of items using in-vector LRU as a building block. We can use in-vector LRU if we use *P* as the size of a set (called *associativity*). However, we observed that a *P*-way set-associative cache using in-vector LRU has a poorer cache hit ratio than exact LRU as a tradeoff for faster processing. A method of improving cache accuracy with a set-associative cache is to increase the associativity. However, increasing the associativity above the processors vector length may attenuate the benefit of faster processing by using in-vector LRU. Also, a set-associative cache based on LRU cannot surpass the precision of the exact LRU since the exact LRU is an extreme case of a set-associative cache with the associativity equal to the cache size.

For achieving both low overhead and high precision, we use our proposed multi-step LRU algorithm in each set. To increase the associativity, each set contains $M$ vectors ($M > 1$), i.e., $M \times P$ items. We can use any $M$, but $M = 2$ or 4 is a reasonable choice to balance the high accuracy and low overhead, as we empirically show later. The first vector contains MRU items and the last ($M$-th) vector contains LRU items. The items in each vector are sorted in the same order, as explained in the previous section. In this section, we use $M = 2$ and $P = 4$ for explanation (hence, the associativity is $2 \times 4 = 8$). Figure 4 illustrates an overview of multi-step LRU. The figure shows only the key of each key-value item to simplify explanation.

When adding a new item $X$ into a set that is already full, we evict the LRU item of the last vector and insert $X$ into the MRU position of the last vector (here, *vector2*) instead of the MRU position of the first vector (*vector1*), i.e., the MRU position of the entire set. This can be easily done by merely applying the put operation of in-vector LRU for the last vector without modifying other vectors. If we apply the exact LRU, we need to place the new item in the MRU position of the first vector and modify all M vectors.

To serve a get query on a key (in the figure, $G$ is the key to search), we scan all vectors from the first to the last by performing the get operation of in-vector LRU. If the query hits in a vector, we return the result after rearranging the items within the vector. Hence, the found item is moved onto the MRU position of the vector but not the MRU position of the first vector. In Figure 4, the first get request moves $G$ to the MRU position of *vector2*. As in the case of the put operation, we modify only one vector and do not need to update other vectors. If the item that matches the query is already at the MRU position of a vector, we swap this matched item with the LRU item of the previous vector. The second get query makes $G$ be included in *vector1* since $G$ is the MRU item in *vector2* when it matches the query. We call this an upgrade. In summary, to manage LRU replacement, a get operation of multi-step LRU either 1) rearranges items within only one vector or 2) swaps two items between two neighboring vectors. The third get query finally makes $G$ the MRU item of the entire set. In general, we need at least ($2M$-1) requests to the same key to place a new item at the MRU position of the entire set.

To help SIMD instructions work efficiently, the current implementation packs all keys and values within a set in a contiguous memory region, as shown in Figure 5.

If a query (e.g. get, put, or delete) for a key comes in, we first calculate a hash value for the key by using a (non-cryptographic) hash function to assign the key to one of the sets. Then we use the above get or put operation for looking up or inserting an item for the specified key. For delete, we conduct a vector comparison to find the key within the set and invalidate the item if found.

As explained above, multi-step LRU is designed to have a low processing cost by limiting the amount of data movements per operation. Surprisingly, multi-step LRU yields a much better hit ratio than exact LRU in addition to its superior execution speed and memory efficiency. In multi-step LRU, only items accessed multiple times in a short period can be upgraded. Hence, if an item is upgraded, it will not be evicted by a long sequence of many items, each of which is accessed only once. Therefore, multi-step LRU achieves a gain similar to sophisticated replacement algorithms that use separated lists for items accessed only once and items accessed twice or more, e.g., 2Q and ARC.

*C. Parallelizing Multi-step LRU*

On today's systems with multiple cores, allowing concurrent accesses from multiple threads is mandatory for fully utilizing system performance. Because our multi-step LRU is based on a set-associative cache, parallelizing it for multi-thread processing is trivial. In a set-associative cache, each query accesses only one set and does not affect other sets. Hence, we can easily implement fine-grained locking by first merely locking a set and releasing the lock at the last of the query. For locking, we need to add additional memory space for a lock within each set, as shown in Figure 5. With such fine-grained locking, multiple queries can be served concurrently if they work on different sets.

IV. EVALUATION

We implemented a caching system with our multi-step LRU using AVX2 instructions and evaluated it on an Intel Xeon E5-2667 v3 processor, which has eight cores running at 3.2 GHz. We implemented the program in C++ using Intel's intrinsics. The system ran under Ubuntu 16.04 Linux distribution. We compiled all the programs using clang++-10.0 with the –O3 option.

To create a query sequence for the evaluations, we used the client emulator of Yahoo Cloud Serving Benchmark (YCSB) [16] for three different distribution patterns called zipfian, latest, and scan. The zipfian (zipf) distribution [17] is suitable for analyzing the cache performance for various types of workloads. In this distribution, the keys are ranked by their frequency of appearing in the sequence, and the relative frequency of the $i$-th item is $1.0/i^{\alpha}$. Here, α is a parameter to determine how the distribution is skewed; larger α results in a more skewed distribution. The original zipfian distribution uses $\alpha=1$, but for web caching, for example, slightly smaller numbers, e.g., around 0.7, are observed in real traces [18]. The latest distribution is similar to zipfian but the distribution is time evolving; it uses the key inserted most recently as the most popular key. The scan distribution accesses a range of keys instead of one key. Based on the use case shown in Figure 1, for each key in the query sequence, we first perform a get operation. If the key is found in the cache, we count it as a cache hit. If the key is not found (i.e. counted as cache miss), we put a new key-value item for the key in the cache. If the cache is already full, we evict one item to insert the new item. To focus on cache replacement performance, we did not do anything with the object in both cache hit and miss cases for all but one experiment.

In all experiments, we used the number of items in the cache to control the cache size regardless of the amount of LRU metadata each algorithm uses. If we use the size of memory space as the cache size, the benefits of multi-step LRU and in-vector LRU in terms of cache hit ratio become larger since they do not use per-item LRU metadata, hence they can cache a larger number of items than other algorithms in the same size of memory space.

### A. Performance of In-vector LRU for small set

We evaluated the execution time and cache hit ratio using a very small cache with a capacity of only four key-value items (both the key and value are 64 bits) to confirm that in-vector LRU can efficiently manage cache replacement for a small cache that can fit within one vector. We tested in-vector LRU implemented using AVX instructions as well as exact LRU based on a link list and GCLOCK. Exact LRU uses two pointers (16 bytes) as LRU metadata per cached item, and GCLOCK uses four bits per item for a reference counter. In contrast, in-vector LRU does not require per-item LRU metadata. In the evaluated implementation, GCLOCK borrowed four bits from the value part of a key-value item, and exact LRU allocated additional memory space separately for the linked list. For cache lookup, both GCLOCK and exact LRU were implemented using SIMD compare instructions, the same as in-vector LRU for fair comparisons.

Figure 6 shows execution times per query for the three replacement algorithms using sequences of one million requests in the zipfian distribution with 10, 20, or 40 distinct keys, i.e., operationcount=1000000 and recordcount=10, 20, or 40 as YCSB workload properties. The algorithms were also tested using two special cases with the cache hit ratios of 0% (shown as *all miss*) and 100% (shown as *all hit*). We observed that in-vector LRU achieved shorter execution time than both GCLOCK and exact LRU for most data distributions. GCLOCK showed the fastest execution time for the all-hit case. This is because GCLOCK merely increments the reference counter for the item to manage replacement when a query hits in the cache as a tradeoff for higher overhead at cache misses. Exact LRU was the slowest among the three algorithms due to the costly bookkeeping of the linked list; GCLOCK is generally faster than exact LRU due to its simpler LRU management.

### B. Performance of Set-associative Cache with Multi-step LRU

Next, we evaluated the performances of cache replacement algorithms with a larger capacity for more realistic cache configurations. We evaluated a 4-way set-associative cache with in-vector LRU (i.e., $P = 4$, $M = 1$) and an 8-way set-associative

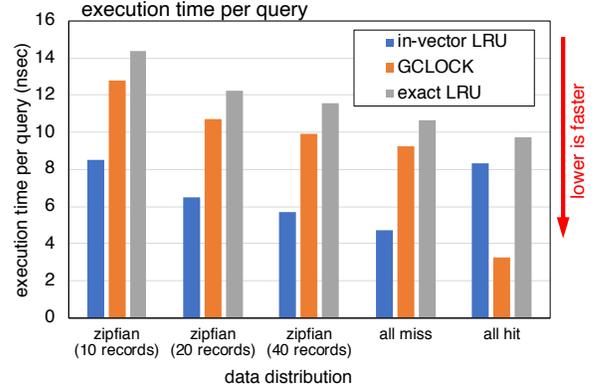

Figure 6. Execution time and number of executed instructions per query for three replacement algorithms for small cache consisting of four key-value items

cache with multi-step LRU (i.e., $P = 4$, $M = 2$). We discussed the effects of the set size (associativity) in detail later. We used MurmurHash3 [19] as the hash function for set-associative caches. Although we use MurmurHash in our implementation, other hash functions, such as xxHash or CityHash, can be used. For comparison, we also tested GCLOCK [9], exact LRU, and ARC [5]. We implemented a cache for exact LRU by using a doubly linked list; the key-value item is added to a hash table for fast lookup and also added to the linked list for LRU management. For the hash table, we used our cuckoo hash table [20] implemented using the same SIMD key lookup component with multi-step LRU to be fair when comparing results. We believe this cuckoo hash table performs reasonably fast; when we use glibc's std::unordered_map instead of our cuckoo hash table, the execution times increase by up to 2.5x.

Figure 7 compares the cache hit ratios with various cache sizes, from 1k to 32M key-value items, for request sequences of 2 billion (for zipfian and latest) or 5 billion (for scan) requests with the 100 million distinct keys. The cache size of 32M key-value items corresponds to 512 MB, hence it cannot fit into the processor's L3 cache memory, the size of which is 20 MB. The trends were almost same for all three data distributions. Among the tested replacement algorithms, ARC achieved the best cache hit ratios with almost all combinations of cache size and distribution. Our multi-step LRU was the second best. GCLOCK's hit ratio was generally lower than those of multi-step LRU and ARC. GCLOCK worked very well for the latest distribution and outperformed multi-step LRU and even ARC when the cache size was large. GCLOCK tends to evict old items first and this characteristic is quite suitable for latest distribution. For all data distributions, exact-LRU showed a poorer cache hit ratio than GCLOCK, multi-step LRU, and ARC. In-vector LRU had a slightly worse cache hit ratio than exact-LRU, hence the worst among all tested replacement algorithms.

Figure 8 illustrates the execution speed of the algorithms in terms of throughput (the number of queries processed divided by the total execution time) in the same experiments. In-vector LRU, which had the worst hit ratio, showed the highest throughput, i.e., fastest execution time. Our multi-step LRU was a close second best. On the other hand, ARC had a much longer execution time than the other algorithms. This is reasonable

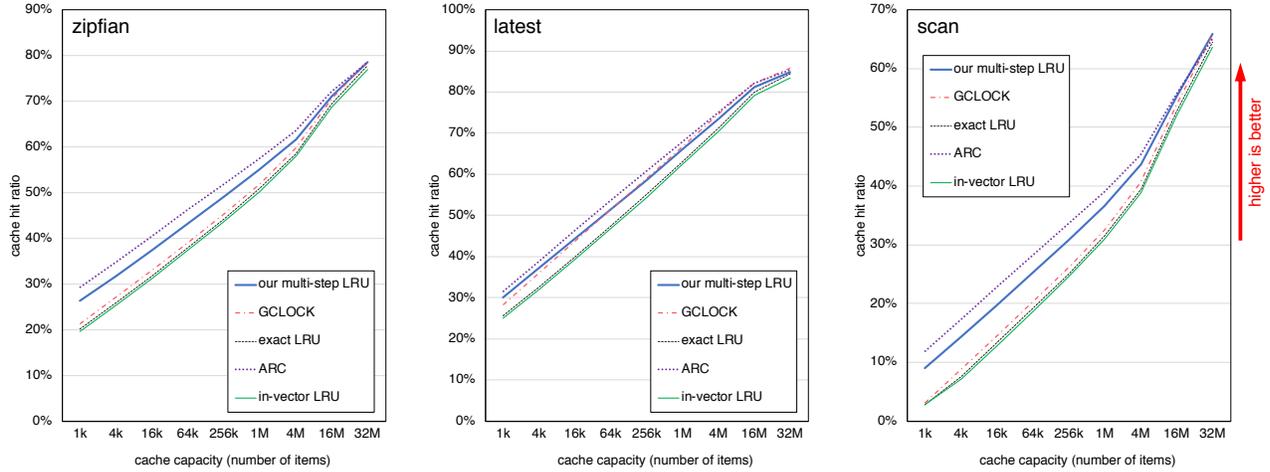

Figure 7. Comparisons of cache hit ratios for query sequences generated by YCSB's client emulator for zipfian, latest, and scan data distributions. X-axis shows cache size in number of items; key range is 1 to 100M; hence, 32M means 32% of all items.

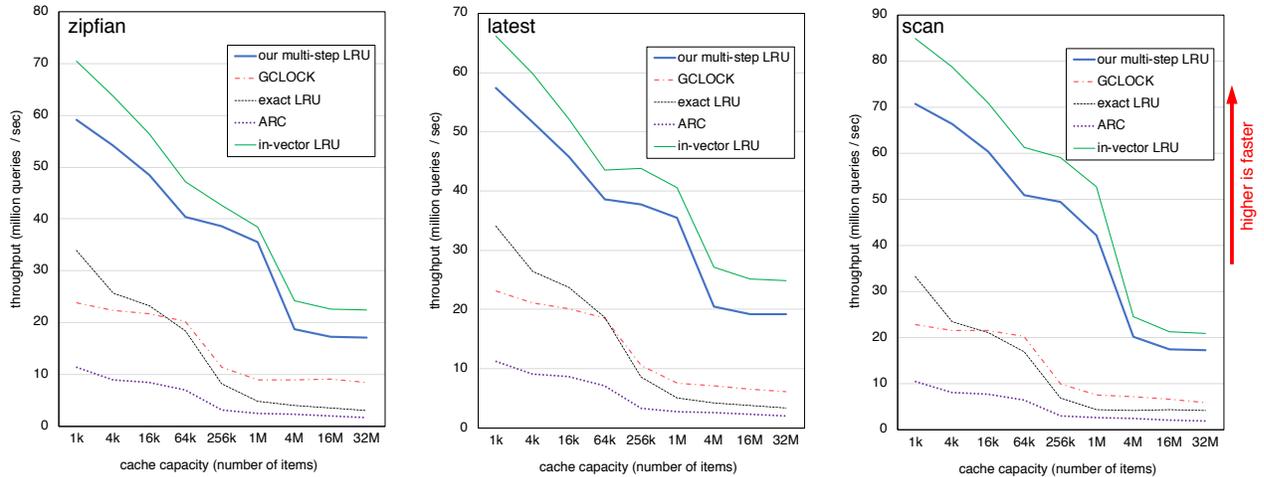

Figure 8. Comparisons of execution times for query sequences that follows zipfian, latest, and scan data distributions. X-axis shows cache size in number of items; key range is 1 to 100M.

because ARC focuses on better precision rather than lower runtime overhead.

The execution times became longer with increasing cache sizes for all algorithms. This is the effect of a processor's cache memory. For small cache size, all cached key-value items and LRU metadata can fit in the L1 or L2 cache memory of the processor. However, a larger cache does not fit within L2 or even L3 cache memory of the processor. Hence, with a larger number of items in a cache, the execution time becomes longer and is more dependent on memory system performance. Since multi-step LRU and in-vector LRU do not use per-item LRU metadata, the performance advantages of these two algorithms are larger with increasing cache sizes; they have very low runtime overhead in both CPU time and memory consumption. ARC and exact LRU use large LRU metadata for doubly linked lists; hence, their performances significantly degrade with larger cache sizes.

To confirm the effect of processor's cache misses on these algorithms, Figure 9 compares the numbers of L3 cache misses and executed instructions (path length) per query as measured by the performance monitor of the processor using Linux's perf command (perf stat -e instructions -e LLC-misses) with two cache sizes, 32M and 64k items. Multi-step LRU and in-vector LRU obviously caused smaller numbers of processor's L3 cache misses than the other three algorithms regardless of the data distribution when the cache size was 32M. When the cache size was 64k, the differences in the L3 cache miss frequencies were not that significant and the absolute numbers of cache misses were small in terms of affecting the execution speed of the algorithms.

The primary reason of the better throughput of multi-step LRU compared to GCLOCK and exact LRU with a small cache size (e.g. 64k cache size) is a shorter path length. Regardless of the cache size, multi-step LRU and in-vector LRU have shorter path lengths than GCLOCK, exact LRU, and ARC because of the simpler replacement algorithms that effectively exploits SIMD instructions.

Since L3 cache misses are one of the bottlenecks that determines overall throughput, we evaluated throughputs while putting more pressure on the memory system by accessing an object after key-value cache accesses. In the above experiments, we obtained the pointer to the object for the key from the cache but not actually touched the object. In this experiment, we actually touched the object, the size of which was 64 bytes. We loaded the entire 64 bytes of the object when a query hit in the cache and wrote the entire 64 bytes when putting the object into the cache after the query did not hit in the cache. Figure 10 shows the throughputs with additional 64-byte memory access per query. Although the throughputs were lower than those without additional memory accesses shown in Figure 8, multi-step LRU yielded higher throughput than GCLOCK and exact LRU, e.g., multi-step LRU showed 37% better throughput than GCLOCK for the cache size of 32M.

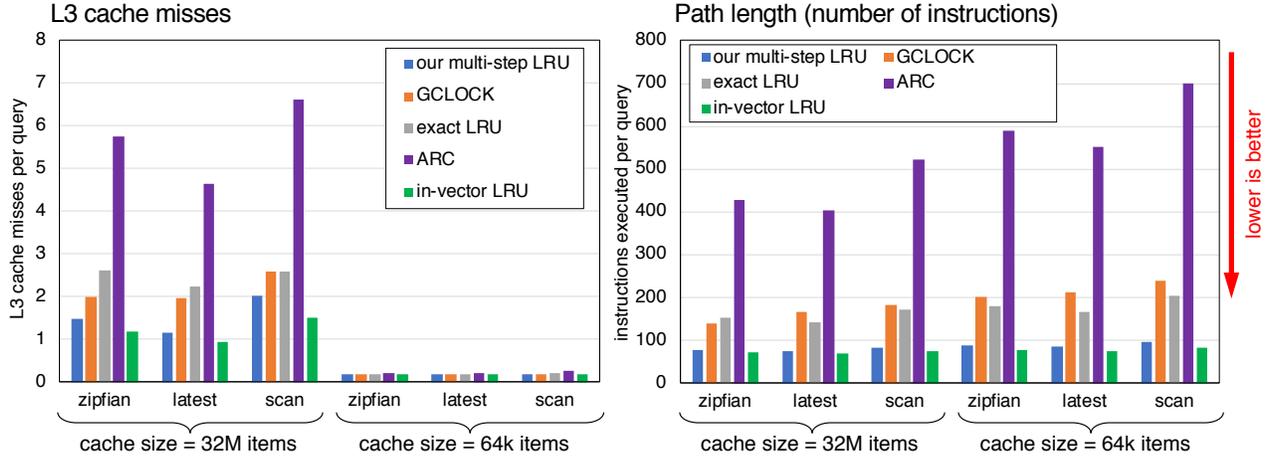

Figure 9. Number of instructions executed per query (path length) and L3 cache misses per query for two cache sizes

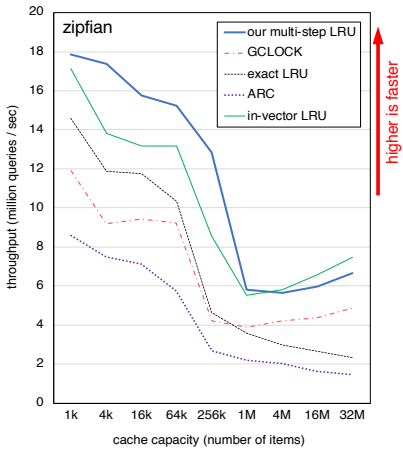

Figure 10. Comparisons of throughput with additional memory accesses

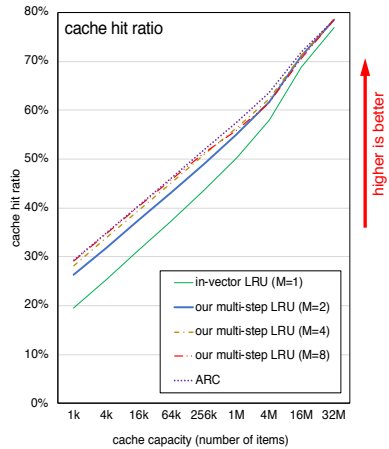

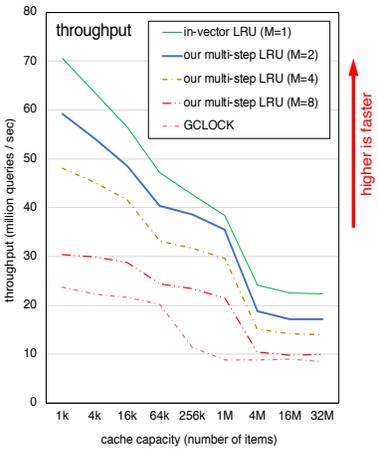

Figure 11. Comparisons of cache hit ratios and execution times with various $M$ parameters for multi-step LRU

These results indicate that multi-step LRU balances the benefits of both high cache precision and high throughput, whereas ARC and in-vector LRU achieves one by sacrificing the other.

*C. Effects of Parameters with Multi-step LRU*

In the previous section, we discussed using multi-step LRU with $M = 2$, i.e., in an 8-way (2 vectors having 4 items each) set-associative cache, as the configuration. We now show how this parameter affects the hit ratio and execution speed.

Figure 11 compares the cache hit ratios and throughputs for the zipfian distribution for multi-step LRU with $M = 2, 4$, and 8. It also shows the results for in-vector LRU ($M = 1$). Note that multi-step LRU becomes identical to in-vector LRU for $M = 1$. Here, $M = 1, 2, 4$, and 8 correspond to 4-way, 8-way, 16-way, and 32-way set associative caches. For comparison, we show ARC's cache hit ratio and GCLOCK's throughput. We can see the cache hit ratio improved with a larger $M$ parameter. When we used $M = 8$, the hit ratio was almost comparable to that of ARC. With increased $M$, the associativity also increased, and a larger associativity generally improves the cache hit ratio. Also, multi-step LRU configured with a larger $M$ can hold more items in the non-last vectors, in which items are protected from being evicted by use-once items. This improvement in cache hit ratio comes at the cost of increased execution time. When we increase the associativity, the overhead for looking up the specified key or an empty space is also increased since we need to check all items in the set. However, even with $M = 8$, the execution time of multi-step LRU is shorter than that of GCLOCK.

Based on these observations, we believe $M = 2$ or 4 is a reasonable choice for balancing speed and precision. Increasing $M$ too much does not significantly improve the cache hit ratio enough to rationalize the degradation in speed.

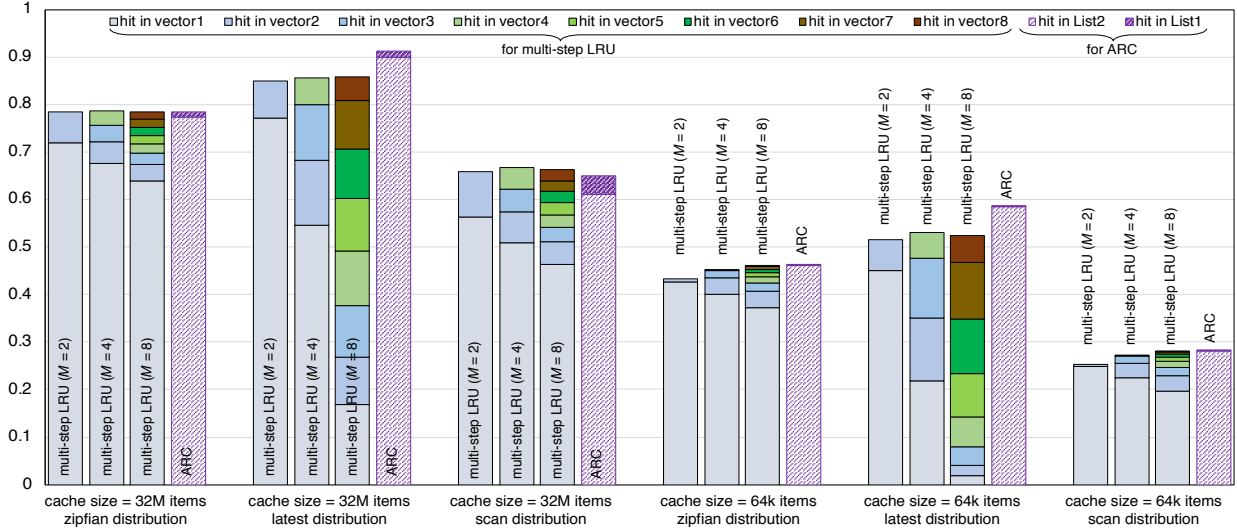

Figure 12. Breakdown of cache hits into locations. First vector contained most frequently used items, as expected.

For further insights into multi-step LRU with different $M$ parameters, Figure 12 shows the breakdown of the cache hits into the location (i.e. vector) in which the queries hit. We also show the breakdown for ARC, which has two lists (list1) for items used only once and another (list2) for items used twice or more. With multi-step LRU, the queries hit in the first vector (vector1) most frequently, as expected. We upgrade only frequently used items by selecting items accessed multiple times in a short period. Hence, in the principle of the algorithm, the first vector should contain the most frequently used items. The results in Figure 12 indicate that our upgrade criteria work well to select frequently used items. With the increased number of vectors, the ratio of query hits in the first vector is reduced. However, even with $M = 8$, more than a half of the queries were hit in the first vector for zipfian and scan. Only for latest distribution, many queries were hit in non-first vectors. This is because hot items are time evolving in the latest distribution and hence one item may not be accessed long enough to be upgraded onto the first vector. For such case, using too large $M$ may not be effective to increase the precision. The breakdown for ARC is consistent with ours; most queries were hit in the list for items used multiple times (list2). ARC adaptively tunes the size ratio between two lists, and the size of list1, for items used only once, is typically much smaller than that of list2, while the sizes of all vectors are uniform and fixed in multi-step LRU. Hence, the breakdown for ARC is significantly skewed toward list2.

Thus far, we discussed testing multi-step LRU with four items per vector ($P = 4$) by assuming 64-bit keys and values (often pointers) on a 256-bit vector register. We now discuss testing multi-step LRU with $P = 8$ by reducing the size of a key and value to 32 bits. Figure 13 compares the cache hit ratios and throughputs. We use $P = 8$ and $M = 2$ for multi-step LRU; hence, the associativity was 16. For in-vector LRU, the associativity was 8 with $P = 8$ and $M = 1$. We compiled all programs as 32-bit binaries to make a pointer fit in 32 bits. Due to the limitation of 32-bit memory space, we used a cache size of up to 4M items. We did not see huge differences in the trends of the cache hit ratios by increasing $P$ from 4 to 8. Our multi-step LRU achieved better cache hit ratios than GCLOCK or exact LRU. Also, there were no significant changes in the relative execution times as well. The fastest was in-vector LRU, and multi-step LRU was a close second best. These results indicate that the high precision and low overhead of multi-step LRU do not depend on a specific data parallelism within one vector ($P$). Hence, multi-step LRU should work on future processors with increased vector length, such as AVX-512.

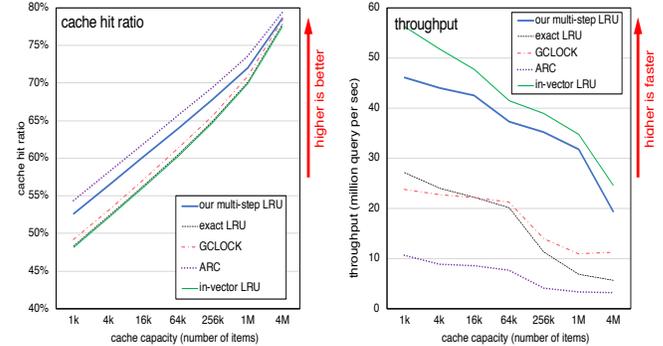

Figure 13. Comparisons of cache hit ratios and throughput of algorithms using longer vector size of $P = 8$ for zipfian distribution.

### D. Scalability with Multiple Cores

We evaluated the performances of the multi-thread implementations of three cache replacement algorithms, multi-step LRU, GCLOCK, and exact LRU. The parallel implementation of multi-step LRU uses fine-grained locking using one lock bit per set. As a memory space, the lock bit uses one byte as shown in Figure 5. If a thread fails to acquire the lock of a set, it waits for another thread to release the lock by spin loop. Since one thread holds a lock only for a short time, using spin loop is much more efficient than using OS-level mutex. For GCLOCK and exact LRU, we used a high-performance concurrent hash map from https://github.com/preshing/junction instead of our cuckoo hash map, which is not thread safe.

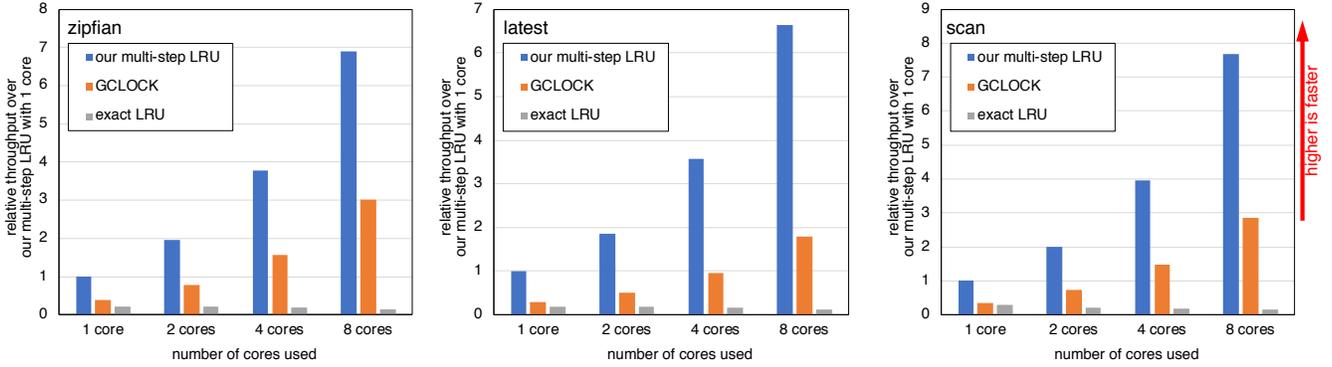

Figure 14. Throughput scalability with increasing number of cores used for multi-step LRU, GCLOCK, and exact LRU

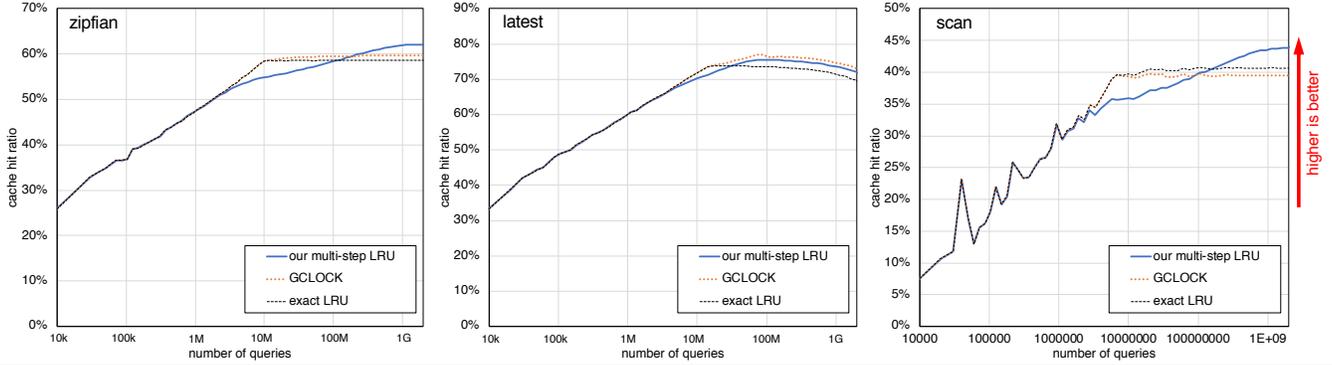

Figure 15. Cache hit ratios during startup time from randomly initialized state. X-axis is number of queries processed in logarithmic scale

Figure 14 compares the relative throughputs of the three algorithms for 1 to 8 cores over the throughput of multi-step LRU with 1 core. The cache size was 32M items. Note that the algorithms with 1 core were slower than of the serial implementations due to locking overheads. Multi-step LRU and GCLOCK showed good scalability with increasing number of cores used with up to 8 cores. With 8 cores, multi-step LRU achieved 6.6x to 7.7x increase in speed and GCLOCK showed 6.3x to 8.0x increase in speed. With 4 cores, the increases were 3.6x to 4.0x for multi-step LRU and 3.4x to 4.0x for GCLOCK. Since both algorithms showed similar scalability, the advantage of multi-step LRU in terms of execution speed remained unchanged from that of serial implementations, as discussed in previous sections. Compared to the other two algorithms, exact LRU does not scale because the lock to guard the LRU linked list becomes the bottleneck of scalability. All queries attempt to update the head of the LRU linked list for updating recently-used item information; hence, it is difficult to improve scalability by applying fine-grained locking.

### E. Warming Up Performance

One possible drawback of multi-step LRU is the potential longer time for warming up, i.e., the time until the cache reaches a steady state. With multi-step LRU, upgraded items are not evicted due to other items accessed only once. This is the main reason of the superior cache hit ratios in steady state. However, once the upgraded items become inactive, e.g., due to a phase change in the access patterns, these now inactive items are not evicted quickly. To evaluate such effects, Figure 15 compares the cache hit ratios during the warmup phase for the three distributions with the cache size of 4M items. We initialized the cache with random keys; hence, all cached items were garbage at first. For all distributions, multi-step LRU required a longer time for the warmup. During a part of the warmup phase, multi-step LRU showed a lower cache hit ratio than exact LRU since a fraction of the cache is filled by already inactive items until they are evicted. Such inactive items may remain in the cache, especially when the cache hit ratio is low, because not so many items are upgraded. GCLOCK and exact LRU do not suffer from long delay before evicting the inactive items. If the workload suffers from this problem, e.g., if the access patterns are frequently changing, we can adaptively switch the algorithm to in-vector LRU, which does exact LRU in each set, during the warmup. For example, we can use a significant drop in the ratio of queries that hit in the first vector as a trigger for switching the algorithm. Note that the case shown in Figure 15 is an extreme case of the phase change where all the hot data become inactive at once. The problem is not that significant if the changes in the access patterns occur gradually. Also, if the initial state of the cache is the empty state instead of full of inactive garbage data, this problem does not occur since multi-step LRU can add new items into the cache if there is an empty slot in any of the vectors.

## V. SUMMARY

We proposed a cache replacement algorithm called multi-step LRU that achieves 1) high memory efficiency, 2) fast processing, and 3) high precision at the same time. Multi-step LRU yields high throughput because it does not use per-item LRU metadata and also efficiently exploit SIMD instructions. It yields a higher cache hit ratio than other widely used LRU or

GCLOCK algorithms by taking both access frequency and access recency of items into account. Since a key-value cache is a key building block of applications that access a huge amount of data with high throughput, multi-step LRU can contribute to real-world workloads by improving cache hit ratios without sacrificing processing efficiency.

## REFERENCES

[1] Memcached - a distributed memory object caching system, https://memcached.org/ Using Redis as an LRU cache, https://redis.io/topics/lru-cache
[2] Using Redis as an LRU cache, https://redis.io/topics/lru-cache
[3] Fan, B., Andersen, D. G., Kaminsky, M. MemC3: compact and concurrent MemCache with dumber caching and smarter hashing. In *Proceedings of the 10th USENIX conference on Networked Systems Design and Implementation* (*NSDI '13*). 2013, 371–384.
[4] Johnson, T., and Shasha, D. 2Q: A Low Overhead High Performance Buffer Management Replacement Algorithm. In *Proceedings of the 20th International Conference on Very Large Data Bases* (*VLDB '94*). 1994, 439–450.
[5] Megiddo, N., and Modha, D. S. ARC: A Self-Tuning, Low Overhead Replacement Cache. In *Proceedings of the 2nd USENIX Conference on File and Storage Technologies* (*FAST '03*). 2003, 115–130.
[6] Wang, R., Wang, Y., Tai, T., Dumitrescu, C. F., Guo, X., Technologies for a least recently used cache replacement policy using vector instructions. US20190042471, July 2019.
[7] Bansal, S. and Modha, D. S., CAR: Clock with Adaptive Replacement. In *Proceedings of the 3rd USENIX Conference on File and Storage Technologies* (*FAST '04*). 2004, 187–200.
[8] Jiang, S., Chen, F., and Zhang, X., CLOCK-Pro: an effective improvement of the CLOCK replacement. In *Proceedings of the annual conference on USENIX Annual Technical Conference* (*USENIX '05*). 2005, 323–336.
[9] Smith, A. J., Sequentiality and prefetching in database systems. *ACM Trans. Database Syst. 3, 3 (Sept. 1978)*. 223–247.
[10] O'Neil, E. J., O'Neil, P. E., and Weikum, G., The LRU-K page replacement algorithm for database disk buffering. In *Proceedings of the ACM SIGMOD International Conference on Management of Data* (*SIGMOD '93*), 1993, 297–306.
[11] Zhou, Y., Philbin, J., and Li, K., The Multi-Queue Replacement Algorithm for Second Level Buffer Caches. In Proceedings of the 2001 USENIX Annual Technical Conference (*USENIX '01*). USENIX Association, USA, 91–104.
[12] Patterson, D. A., and Hennessy, J. L., *Computer Organization and Design: The Hardware/Software Interface.* Morgan Kaufmann Publishers Inc. 2013.
[13] Intel corp., Intel C++ Compiler Developer Guide and Reference.
[14] Inoue, H., Komatsu, H., Nakatani, T., Accelerating UTF-8 Decoding Using SIMD Instructions (in Japanese), *Information Processing Society of Japan Transactions on Programming* **1** (2), 2008, 1–8.
[15] Stepanov, A. A., Gangolli, A. R., Rose, D. E., Ernst, R. J., and Oberoi, P. S., SIMD-based decoding of posting lists. In *Proceedings of the 20th ACM international conference on Information and knowledge management* (*CIKM '11*). 2011, 317–326.
[16] Cooper, B. F., Silberstein, A., Tam, E., Ramakrishnan, R., and Sears, R., Benchmarking cloud serving systems with YCSB. In *Proceedings of the 1st ACM symposium on Cloud computing (SoCC '10)*. 2010. 143–154.
[17] Zipf, G. K., Relative Frequency as a Determinant of Phonetic Change. *Harvard Studies in Classical Philology* **40,** 1929, 1–95.
[18] Breslau, L., Cao, P., Fan, L., Phillips, G., and Shenker, S., Web caching and Zipf-like distributions: evidence and implications. In *Proceedings of IEEE INFOCOM '99. Conference on Computer Communications*. 1999, 126–134.
[19] Appleby, A., MurmurHash3, https://github.com/aappleby/smhasher/
[20] Pagh, R., and Rodler, F. F., Cuckoo hashing. *J. Algorithms* **51**, 2 (May 2004), 122–144.

# Appendix

```c
const int patternTable[] = {0,1, 2,3, 4,5, 6,7,  // pattern to move the first item to MRU (do nothing)
                            2,3, 0,1, 4,5, 6,7,  // pattern to move the second item to MRU
                            4,5, 0,1, 2,3, 6,7,  // pattern to move the third item to MRU
                            6,7, 0,1, 2,3, 4,5}; // pattern to move the last (LRU) item to MRU

int64 get(int64 *pKeys, int64 *pVals, int64 keyToSearch) {
  // 1) Load P keys from memory into a vector register.
  __m256i vKeys = _mm256_load_si256(pKeys);

  // 2) Check P keys against the query using SIMD compare.
  // `bitmask` shows hit or not for each comparison. `pos` shows the position of the hit.
  int bitmask = _mm256_movemask_pd(_mm256_cmpeq_epi64(_mm256_set1_epi64x(keyToSearch), vKeys));
  int pos = 31 - _lzcnt_u32(bitmask);

  // 3) If no key hits, return here (a cache miss).
  if (pos < 0) return CACHE_MISS;

  // 4) Look up the in-memory pattern table using the position of the matched key (pos) as the index.
  const __m256i pattern = _mm256_load_si256(((__m256i*)patternTable) + pos);

  // 5) Move the hit key into the MRU position by a permutation and store back the rearranged keys.
  vKeys = _mm256_permutevar8x32_epi32(vKeys, pattern);
  _mm256_store_si256(pKeys, vKeys);

  // 6) Load values into a vector register, rearrange them using the same pattern, and store back.
  __m256i vVals = _mm256_load_si256(pVals);
  _mm256_store_si256(pVals, _mm256_permutevar8x32_epi32(vVals, pattern));

  // 7) Return the result (a cache hit). The result is always stored in the first element.
  return pVals[0];
}

void put(int64 *pKeys, int64 *pVals, int64 newKey, int64 newVal) {
  // 1) Load P keys into a vector register.
  __m256i vKeys = _mm256_load_si256(pKeys);

  // 2) move the LRU key into the MRU position.
  const __m256i pattern = _mm256_load_si256(((__m256i*)patternTable) + P - 1);
  vKeys = _mm256_permutevar8x32_epi32(vKeys, pattern);

  // 3) Replace the LRU key with the new key and then store back them into memory.
  vKeys = _mm256_insert_epi64(vKeys, newKey, 0 /* position */);
  _mm256_store_si256(pKeys, vKeys);

  // 4) do step 1 to 3 for values
  ... (omitted)
}
```

**Figure.** Pseudo code of in-vector LRU using Intel's intrinsics for AVX instruction